\documentclass{svmult}

\usepackage{amsmath,amssymb}
\usepackage{makeidx}         
\usepackage{graphicx}        
\usepackage{multicol}        
\usepackage[bottom]{footmisc}

\makeindex             

\begin{document}

\title*{Three-frame Algorithm of Car Path Reconstruction from Airborne Traffic Data}
\titlerunning{Three-frame Algorithm of Car Path Reconstruction}
\author{Ihor Lubashevsky\inst{1}\and
        Namik Gusein-zade\inst{1}\and
        Dmitry Klochkov\inst{1}\and
        Sergey~Zuev\inst{2}
}
\authorrunning{I. Lubashevsky\and D. Klochkov\and N. Gusein-zade\and S. Zuev}
\institute{A.M.~Prokhorov General Physics Institute, Russian Academy of
           Sciences, Vavilov str. 38, Moscow, 119991, Russia,
          e-mail~(IL):~\texttt{ialub@fpl.gpi.ru} \and
          German Aerospace Center, Optimal Information Systems, Rutherfordstrasse 2, D-12489
          Berlin, Germany}
%
%
\maketitle

\abstract {The airborne traffic monitoring system forms a novel technology of
detecting vehicle motion. An optical digital camera located on an airborne
platform produces a series of images which then are processed to recognized the
fixed vehicles. In this way the video data are converted into the time sequence
of frames containing the vehicle coordinates. In the present work a three-frame
algorithm is developed to identify the succeeding vehicle positions. It is
based on finding the neighboring points in the frame sequence characterized by
minimal acceleration. To verify and optimize the developed algorithm a
``Virtual Road'' simulator was created. Finally available empirical data are
analyzed using the created algorithm.}

\subsection*{Introduction} \label{sec:1}

The traditional techniques of traffic flow measurements are based on local
detectors mounted at fixed places of a road network. Such detectors provide
adequate information about traffic flow only for road fragments of a rather
simple structure. However in cities there are a large number of complex road
intersections, where vehicles moving on different branches interact strongly
with one another. To measure appropriately traffic flow features on such ``hot
areas'' the information supplied by the stationary detectors should be
complemented at least with the detailed information about the spatial structure
of traffic streams on these hot areas.
Currently the novel technology of airborne monitoring and detecting the main
characteristics of traffic flow is under development.  The German Aerospace
Center (DLR) with its department of Transportation Studies and department of
Optical Information Systems has developed an airborne traffic monitoring system
which was successfully used during the Soccer World Championship 2007 in
Germany \cite{1,2,3}. In particular, these and some other DLR airborne datasets
are used in the present paper.
This system is based on an optical sensor placed on an airborne platform. The
optical sensor produces a series of images of area under. The empirical
airborne traffic data obtained by recognizing these images form a set of the
frames with individual vehicle coordinates. To use these data and to calculate
the main parameters of traffic flow trajectories of individual vehicle motion
have to be reconstructed. In this way it becomes possible to measure the
traffic flow rate, the mean velocity of vehicles, and their density
simultaneously.

\subsection*{Vehicle Trajectory Reconstruction}\label{sec:2}

The airborne traffic data under consideration are the time series $\mathcal{F}$
of frames $\{F_t\}$ with the coordinates of recognized vehicles,
%
$
 \mathcal{F} = \{F_{t} = \{x_\alpha  ,y_\alpha \}\}
$.
%
The problem is to construct a collection of trajectories
%
$
 \mathcal{P} = \{ x(t),y(t)\} _{t }
$
%
passing through the points $\{\{x_\alpha  ,y_\alpha \}\}$ at the corresponding
time moments $\{t=n\tau\}$, where $n$ is integer and $\tau$ is the time span
between successive frames. Some constraints should be imposed on the vehicle
trajectories $\mathcal{P}$ to make them smooth. There are several ways to do
this, in particular, to bound the ``acceleration'' (the used approach)
\begin{subequations}\label{eq:3}
\begin{align}
\label{eq:3a}
    \sqrt {\left( {\frac{{d^2 x_\alpha  }} {{dt^2 }}} \right)^2  + \left(
    {\frac{{d^2 y_\alpha  }} {{dt^2 }}} \right)^2 } & \leq a_{\text{max} }\\
 \intertext{or to bound the ``velocity''}
\label{eq:3b}
    \sqrt {\left( {\frac{{dx_\alpha  }} {{dt}}} \right)^2  + \left(
    {\frac{{dy_\alpha  }} {{dt}}} \right)^2 } & \leq v_\text{max}\,.
\end{align}
\end{subequations}
We note that the acceleration and velocity thresholds, $a_\text{max}$ and
$v_\text{max}$, are not the real characteristics of car motion but internal
algorithm parameters whose choice is determined by its efficiency being
highest.

\begin{figure}[t]
\begin{center}
\includegraphics[width=0.7\textwidth]{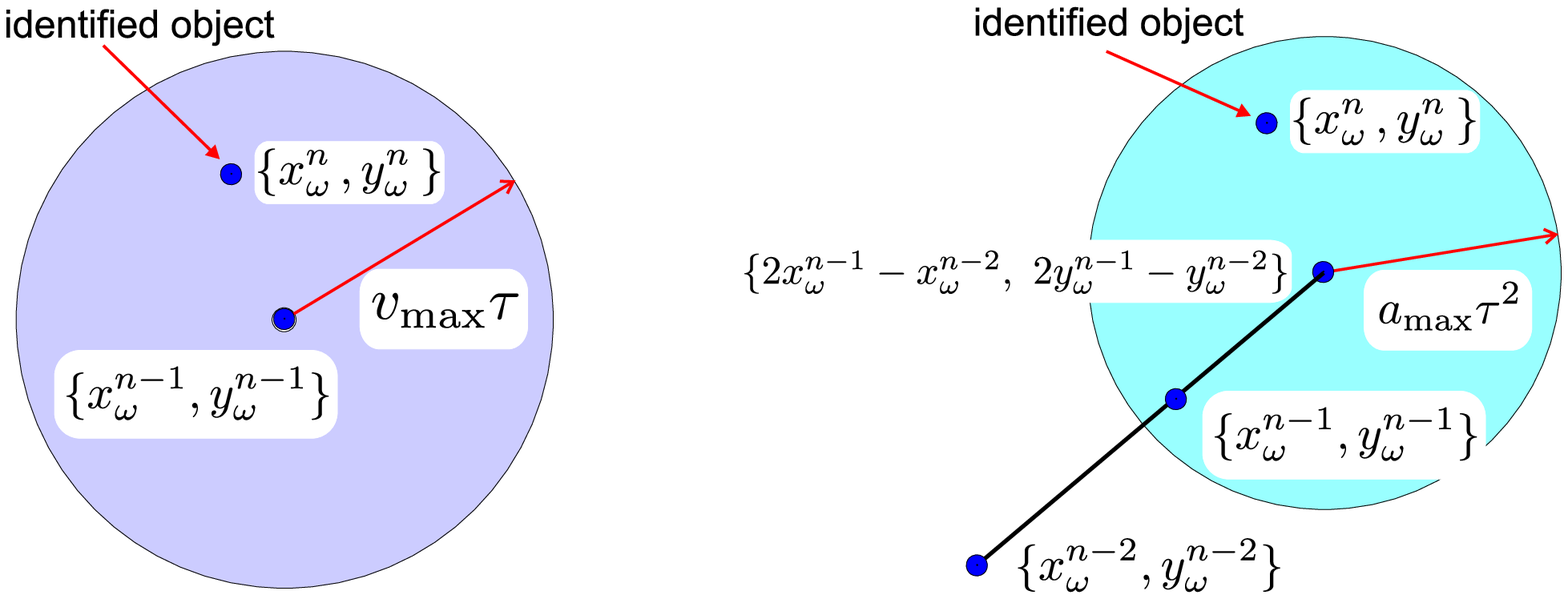}
\end{center}
\caption{Illustration of the car identification algorithms. Within the
two-frame speed limitation algorithm (left fragment) new objects $\{ x^n ,y^n
\}$ are sought in the circle of radius $r = v_{{\text{max}}} \tau$ centered at
the point $\{ x_\omega ^{n - 1} ,y_\omega ^{n - 1} \} $. In the three-frame
algorithm bounding acceleration (right fragment) new objects $\{ x^n ,y^n \}$
are sought in the circle of radius $r = a_{{\text{max}}} \tau ^2 $ centered at
the point $\{ 2x_\omega ^{n - 1}  - x_\omega ^{n - 2} ,\,\,2y_\omega ^{n - 1} -
y_\omega ^{n - 2} \}$.} \label{Fig1}
\end{figure}

To explain the crux of the algorithm implementing these constrains let us
assume that the preceding frames $\{\ldots,F_{n - 2},\;F_{n - 1}\}$ have been
analyzed and the vehicle trajectories are reconstructed at the previous time
moments $\{\ldots,n - 2,\;n - 1\}$. Then a car $\{x_\alpha ^n ,y_\alpha ^n \}$
in the frame $F_n $ will be incorporated into the trajectory $\omega  = \cup_m
\{ x_\omega ^m ,y_\omega ^m \}$ if
\begin{equation}\label{eq:acc}
\sqrt {(x_\omega ^{n - 2}  - 2x_\omega ^{n - 1}  + x_\alpha ^n )^2  + (y_\omega
^{n - 2}  - 2y_\omega ^{n - 1}  + y_\alpha ^n )^2 }  \leq a_{{\text{max}}}
\,\tau ^2\,.
\end{equation}
Figure~\ref{Fig1} illustrates the given three-frame algorithm based on
(\ref{eq:3a}) as well as the relative two-frame algorithm based on
(\ref{eq:3b}) mentioned for comparison only.

\begin{figure}[t]
\begin{center}
\includegraphics[width=0.8\textwidth]{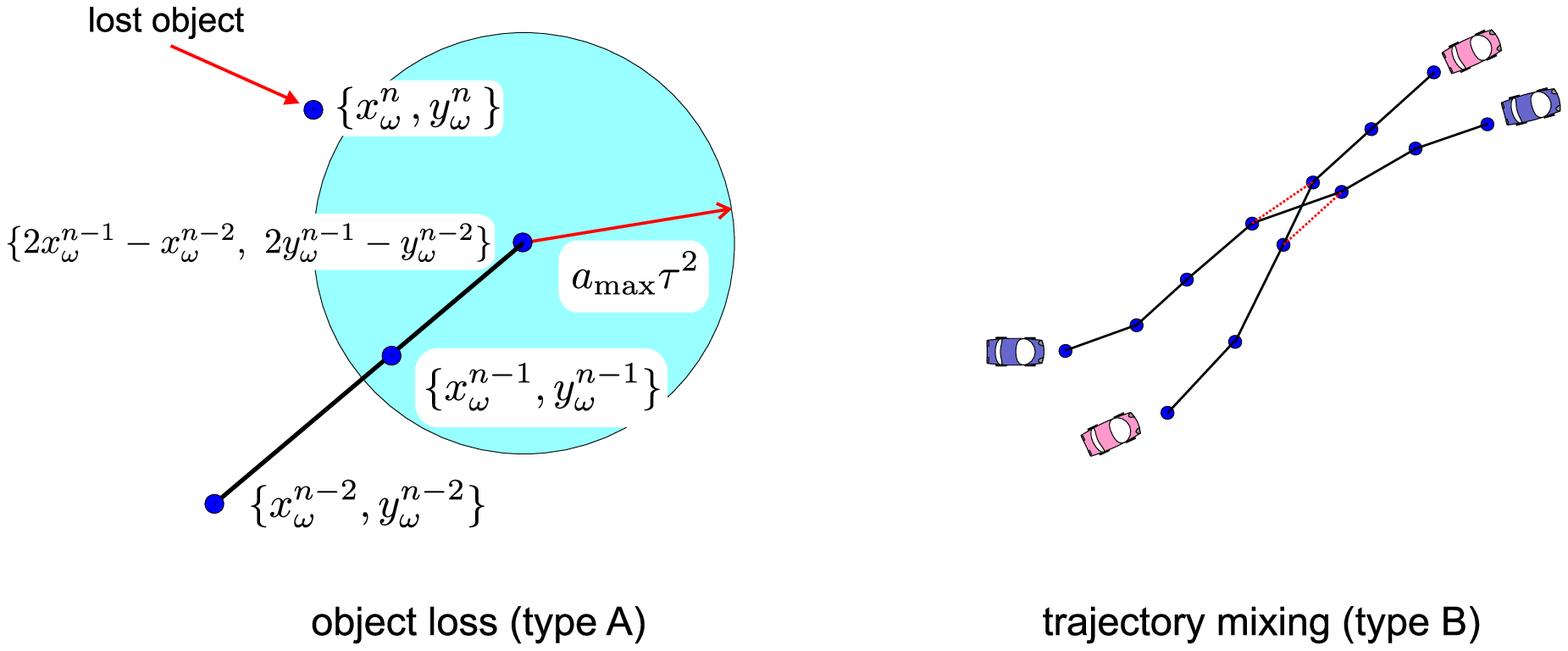}
\end{center}
\caption{The main errors of the car identification algorithm.}
\label{Fig2}
\end{figure}

Let us discuss the possible errors of the car identification algorithm caused
by errors in the car coordinate measurements. There are two main types of the
latter ones; the errors of individual vehicle positions, $\xi_1$, and the
errors in the frame reference to GPS, $\xi_2$. According to the pilot airborne
monitoring system (Institute for Traffic Research, DLR, Berlin) \cite{1,2} the
error values can be estimated as $\xi_1\sim 0.5$~m and $\xi_2\sim 1$--2~m. The
caused errors of the car identification algorithm, i.e. the object loss (type
A) and the trajectory mixing (type B) illustrated in Fig.~\ref{Fig2} will be
quantified in units of
\begin{equation*}
A(B) = \frac{\text{The number of lost(mixed) objects}} {\text{Total number of
objects}} \cdot 100\%
\end{equation*}

\begin{figure}[t]
\begin{center}
\includegraphics[width=0.9\textwidth]{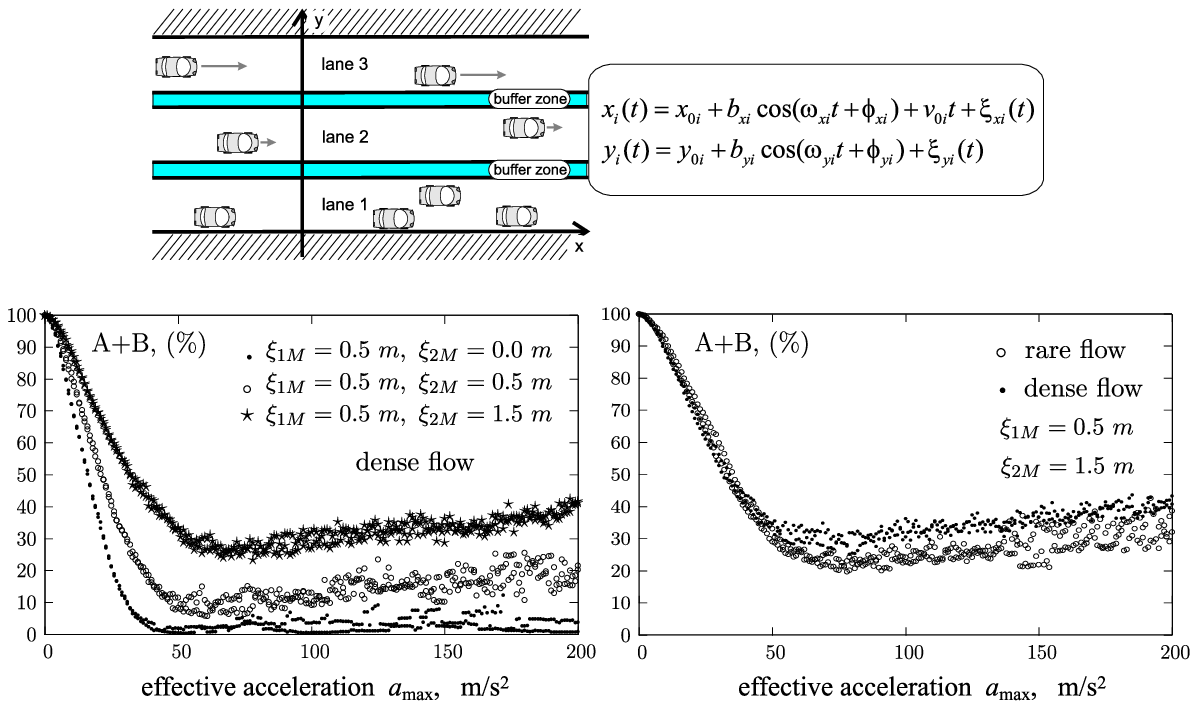}
\end{center}
\caption{Airborne data simulator, ``Virtual Road'', used to verify the vehicle
identification algorithm (upper fragment) and the corresponding results of the
car identification routine (lower fragment). The subscript $M$ at the noise
components $\xi_{iM}$ stands for their amplitudes.} \label{Fig3}
\end{figure}

To analyze the feasibilities of the identification algorithm, first, the
following kinematic simulator has been used (Fig.~\ref{Fig3}, upper fragment).
The motion of cars is specified by the shown equations where the parameters
$x_{0i}$, $y_{0i}$, $b_{xi}$ etc. are random and some addition constraints are
imposed to prevent the collisions. The noise components $\xi_{x(y)i}
=(\xi_1+\xi_2)_{x(y)i}$ imitate the errors in the measurements of the car
coordinates $x_i$ and $y_i$ at the time moments $t=n\tau$. The system
parameters and the number of cars are specified so that the virtual car
ensemble imitate the real characteristics of traffic flow, at least
semiquantitatively. The lower fragment in Fig.~\ref{Fig3} exhibits the net
value of the identification errors, $A+B$, vs. the acceleration threshold
$a_\text{max}$. As could be expected, the present results demonstrate the fact
that the car identification algorithm attains its maximal efficiency when
\begin{equation}\label{eq:4}
a_\text{max}  = \frac{4(\xi_{1M}  + \xi_{2M} )}{\tau ^2 }\,,
\end{equation}
and, in particular, for $\xi_{1M} =0.5$~m and $\xi _2  = 1.5$~m the optimal
acceleration threshold is about $a_\text{max}\sim 50$~m/s$^2$.

\begin{figure}
\centerline{
\includegraphics[width=0.8\textwidth]{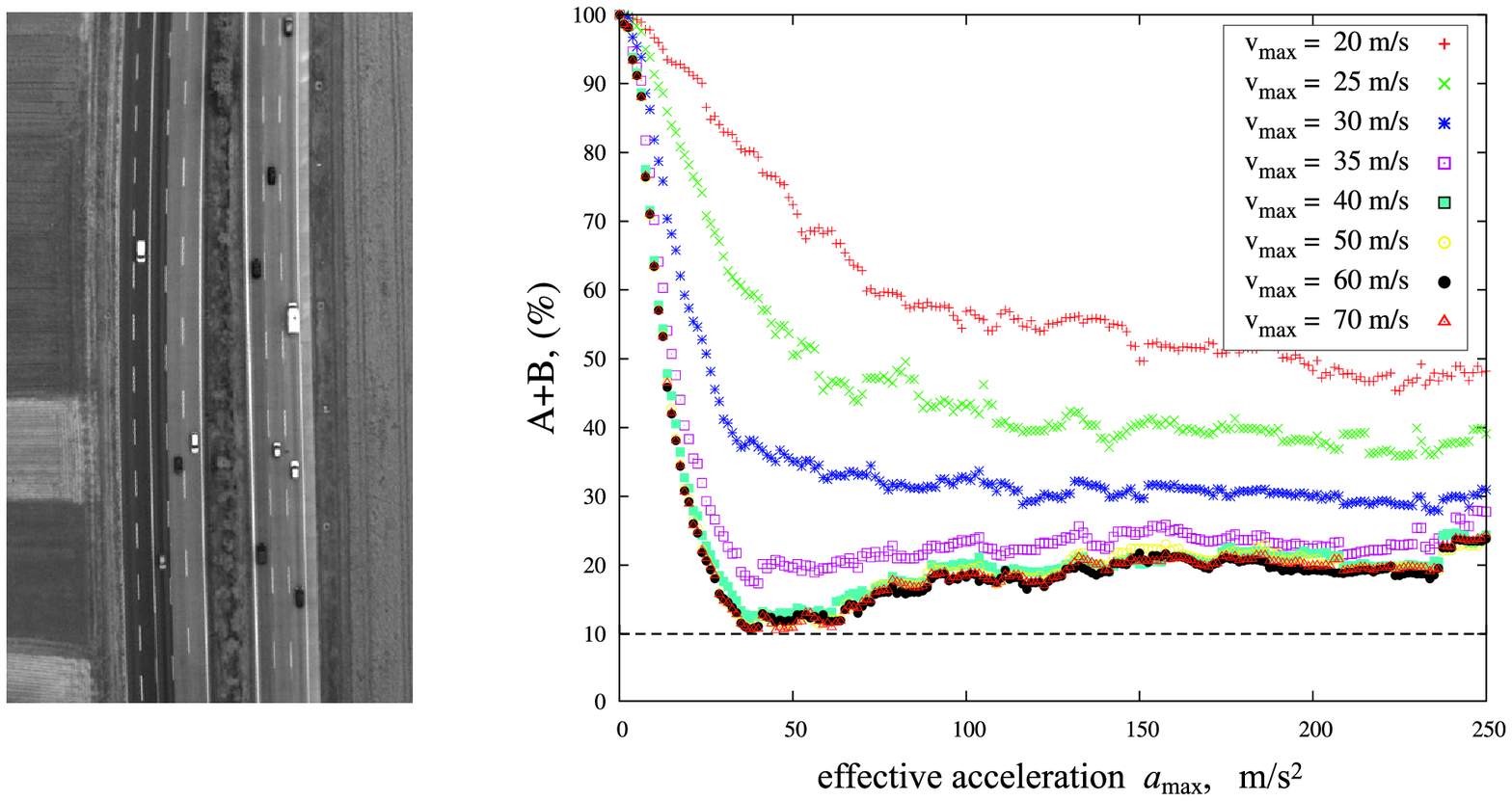}}
\caption{The airborne data collected during the Soccer Championship 2006 in
Stuttgart. Typical example of the analyzed video data (left fragment) and the
results of car identification routine vs. the acceleration threshold (right
fragment).} \label{Fig4}
\end{figure}
Second, the developed algorithm has been verified using the available airborne
traffic data collected during Soccer Championship 2006 in Stuttgart within the
DLR project ``Soccer'' \cite{3}. These data undergone manual processing, so
initially we had a collection of vehicle trajectories regarded as real keeping
in mind the human abilities. The vehicle identification algorithm has given the
result depicted in Fig.~\ref{Fig4}. For these data, again, the optimal value of
the acceleration threshold meets relation~\eqref{eq:4}. Figure~\ref{Fig4}
presents also the efficiency of the car identification algorithm for different
values of the velocity threshold $v_\text{max}$. As seen the imposition of the
additional speed limitation constraint reduces the algorithm efficiency. In
other words, constraints~\eqref{eq:3} interfere with each other and, thus,
should be used separately. As should be expected the measurement errors and the
optimal acceleration threshold are again related by expression~\eqref{eq:4}.



\begin{figure}[t]
\begin{center}
\includegraphics[width=\textwidth]{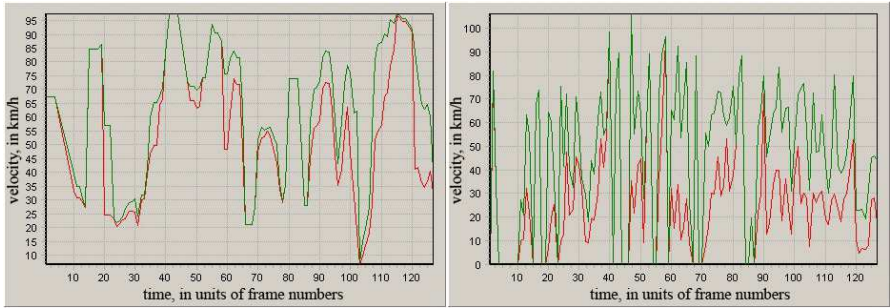}
\end{center}
\caption{Time-mean speed (green) and space-mean speed (red) reconstructed using
the smoothed vehicle trajectories (left fragment) and the same data obtained
using the neighboring points directly. The airborne data of pilot flights in
Berlin, 2004, the LUMOS project.} \label{Fig5}
\end{figure}
\begin{figure}
\begin{center}
\includegraphics[width=0.8\textwidth]{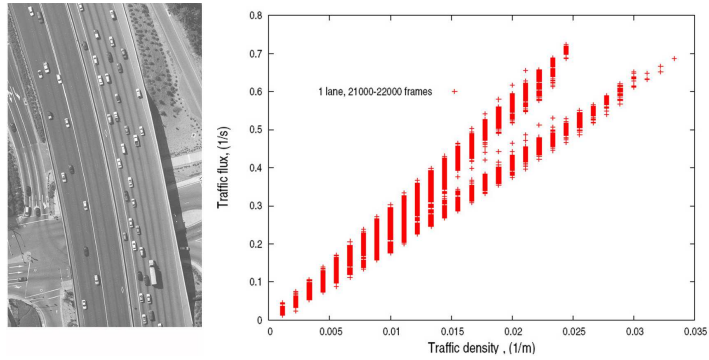}
\end{center}
\caption{Fundamental diagram (right fragment) and an example of the analyzed
video data (left fragment). The NGSIM ``First Prototype'' dataset \cite{4}.}
\label{Fig6}
\end{figure}

\subsection*{Mean Velocity of Traffic Flow and Fundamental Diagram}

Because of the considerable errors in the vehicle coordinates the found
successive positions of a car cannot be used directly to calculate its
velocity. However, the reconstructed trajectories enable one to overcome this
problem by fitting a rather smooth trajectory to the found data. In this way we
have analyzed the empirical data collected during pilot flights of an airplane
in Berlin, 2004 (DLR, LUMOS project). The mean velocity of traffic flow was
calculated, first, by smoothing the reconstructed vehicle trajectories via the
Savitzky-Golay filter and, then, averaging the velocities over the observed car
ensemble. The results are illustrated in Fig.~\ref{Fig5}. As seen the time
pattern of traffic flow speed obtained by smoothing the reconstructed vehicle
trajectories gives much more adequate description of traffic flow.

Using the developed technique we have analyzed also the NGSIM ``First
Prototype'' dataset consisting of vehicle trajectories on a half-mile section
of Interstate 80 in Emeryville, California, for one half of hour \cite{4}. The
system portrait on the phase plane ``traffic flow rate -- car density'' was
drawn using the mean car velocity calculated via the developed technique and
car density obtained by just counting the detected objects (Fig.~\ref{Fig6}).
No widely scatted states are visible, instead, branching is fixed. This result
again poses the question about the necessity to single out the vehicle flow
discreteness in analyzing the states of traffic flow \cite{we}. It should be
noted that the developed approach depresses this discreteness effect.

\subsection*{Conclusion}

We have developed the three-frame-algorithm of reconstructing vehicle
trajectories from airborne traffic data, which evaluates the proximity of
objects in the neighboring frames in terms of acceleration. The acceleration
threshold $a_\text{max}$  is determined by the car measurement accuracy. Using
the ``Virtual Road'' simulator and the dataset collected within the project
``Soccer'' (DLR, Stuttgart, 2006) the value of $a_\text{max}\sim
40$--50~m/s$^2$ is found and the possibility of decreasing the identification
error down to 5--30~\% is shown.

Using one of the NGSIM datasets it has been demonstrated that for the observed
congested traffic the fundamental diagram splits into two branches when the car
density exceeds some value rather than exhibits the widely scatted states. This
result again poses a question whether the discreteness of traffic flow could be
responsible for the appearance of widely scatted states on the fundamental
diagrams. Naturally, the unambiguous answer requires special and detailed
investigation.

\begin{acknowledgement}
This work was supported in part by INTAS project 04-78-7185, DFG project MA
1508/8-1, and RFBR grants 06-01-04005, 05-01-00723, and 05-07-90248.
\end{acknowledgement}



\end{document}